\documentclass[fleqn,10pt]{wlscirep}

\title{Semiquantum key distribution with secure delegated quantum computation}

\author[1,2,*]{Qin Li}
\author[2]{Wai Hong Chan}
\author[3]{Shengyu Zhang}

\affil[1]{College of Information Engineering, Xiangtan
University, Xiangtan 411105, China}
\affil[2]{Department of Mathematics and Information Technology, The Hong Kong Institute of Education, Hong Kong}
\affil[3]{Department of Computer Science and Engineering, The Chinese University of Hong Kong, Hong Kong}
\affil[*]{liqin@xtu.edu.cn}

\begin{abstract}
Semiquantum key distribution allows a quantum party to share a random key with a ``classical" party who only can prepare and measure qubits in the computational basis or reorder some qubits when he has access to a quantum channel. In this work, we present a protocol where a secret key can be established between a quantum user and an almost classical user who only needs the quantum ability to access quantum channels, by securely delegating quantum computation to a quantum server. We show the proposed protocol is robust even when the delegated quantum server is a powerful adversary, and is experimentally feasible with current technology. As one party of our protocol is the most quantum-resource efficient, it can be more practical and significantly widen the applicability scope of quantum key distribution.
\end{abstract}

\begin{document}

\flushbottom
\maketitle
%
%
\thispagestyle{empty}

\section*{Introduction}
Conventionally quantum key distribution requires that two remote parties (usually called Alice and Bob) should have somewhat quantum capabilities to establish a shared key, such as the ability of preparing and measuring qubits in different bases. However, not all of the users own enough quantum resources or have equal quantum technologies in reality. Moreover, a protocol sometimes may not need to be completely quantum to obtain a significant advantage over all its classical counterparts. Based on these two points, not fully quantum key distribution was first introduced by Boyer et al. \cite{BKM:SQKD:PRL07} where secure key distribution becomes possible when one party Alice is quantum, yet the other party Bob has only ``classical'' capabilities, which means someone is limited to perform the following four operations: (1) prepare qubits in the computational basis $\{|0\rangle,|1\rangle\}$, (2) measure qubits in the computational basis $\{|0\rangle,|1\rangle\}$, (3) reorder qubits, and (4) access quantum channels. The party Bob with such limitation is customarily called ``classical" Bob, and this kind of protocol is termed as ``quantum key distribution with classical Bob" or ``semiquantum key distribution (SQKD)".

The first SQKD protocol was proposed in 2007 by using four quantum states, each of which is randomly prepared in the rectilinear or diagonal basis \cite{BKM:SQKD:PRL07}. The idea was extended further and two similar protocols were presented in Ref. \citenum{BGKM:SQKD:PRA09}. One is based on measurement and
the other is based on randomization. Almost simultaneously, Ref. \citenum{zou2009semiquantum} showed the SQKD protocol in Ref. \citenum{BKM:SQKD:PRL07} can be simplified by employing less than four quantum states and proposed five different SQKD protocols using three quantum states, two quantum states, and one quantum state, respectively. In 2011, a more efficient SQKD protocol was proposed based on entangled states \cite{jian2011semiquantum}, where the qubit efficiency is improved to $50\%$, compared with $25\%$ of the protocol in Ref. \citenum{BKM:SQKD:PRL07}. Recently, Ref. \citenum{zou2015semiquantum} proposed an SQKD protocol in which the ``classical'' party does not need the measurement capability, and just needs preparing, sending and reordering qubits. All these SQKD protocols generally assume the existence of an authenticated classical channel, which can be removed by preshareing a master secret key between the communicants \cite{yu2014authenticated}. Furthermore, several multiuser SQKD protocols were put forward \cite{lu2008quantum,zhang2009quantum,krawec2014mediated}. The protocol in Ref. \citenum{lu2008quantum} allows quantum Alice to share a key with several ``classical" participants Bob$_1$, Bob$_2$, $\cdots$, Bob$_n$. The protocols in Refs. \citenum{zhang2009quantum,krawec2014mediated} allow two ``classical" participants to generate a shared key with the aid of an untrusted quantum server. In addition, other semiquantum cryptographic issues beyond SQKD have also been studied to some extent \cite{li2013quantum,wang2012semiquantum,li2010semiquantum,nie2013semi,zou2014three}.

However, in the all above-mentioned semiquantum cryptographic protocols, so-called ``classical" users are not really classical since they still need some quantum ability of preparing and measuring qubits in the computational basis, or quantum memory to reorder qubits. That means they still require corresponding quantum devices to perform certain operations. Then we give a protocol for a nearly classical party Bob who does not possess any quantum device except those necessary for accessing quantum channels to share a key with quantum Alice by the delegation of quantum computations (DQC). In other words, such Bob does not need to implement operations (1), (2), and (3), and only requires the ability to perform the operation (4). But in the presented protocol, there may be not only an independent eavesdropper Eve attempting to obtain some information about the shared key, the delegated server Charlie also may become a powerful adversary. Note that the delegated server can be Alice if she can implement some complicated quantum operations that Charlie needs. But in this case, Charlie becomes a trusted quantum server and Eve is the only attacker. It is obvious that any attack that Eve tries to do may be absorbed into the untrusted Charlie's attack. Therefore, we will show the proposed SQKD protocol is robust like typical SQKD protocols even when Charlie is malicious.

\section*{The review of DQC}
In order to design the new SQKD protocol, we will utilize the technique of DQC. It is quite useful and attracts much attention recently \cite{fisher2014quantum,Broadbent2015delegating,Childs:BQC:QIC05,AS:BQC:IJQIC06,BFK:UBQC:FOCS09,morimae2013secure,sheng2015deterministic,BEBFZP:DBQC:S12,MF:BQC:NC12,Morimae:BQC:PRL12,
DKL:BQC:PRL12,MF:BQC:PRA13,SKM:BQC:PRA13,GMMR:BQC:PRL13,MPFF:BQC:PRL13,li2014triple,HMveri2015,dupuis2012actively,BGS:BQC:AC13} since it can enable users with limited quantum power to perform quantum computation while still keeping users' data private. For instance, Broadbent et al. presented the first universal protocol for DQC where the client only needs to be able to prepare single-qubite states, and a double-server protocol where the client can be totally classical with the assumption that two servers should be non-communicating \cite{BFK:UBQC:FOCS09}. Morimae and Fujii utilized the one-way hashing distillation model to skillfully realize entanglement distillation for the double-server protocol in Ref. \citenum{BFK:UBQC:FOCS09} and gave a modified protocol \cite{morimae2013secure} to adapt to the case after entanglement distillation. Then Sheng et al. employed hyperentanglement to give a much simpler way to achieve secure distillation for the same double-server protocol with the success probability of 100\% \cite{sheng2015deterministic}, which will greatly increase the practical application for the protocol in a noisy quantum channel. Recently, Li et al. removes the demanding requirement that two servers cannot communicate with each other in double-server BQC protcols and gave a more practical DQC protocol \cite{li2014triple}. Although many protocols have been proposed, there are mainly three kinds of methods to achieve DQC, including applying universal quantum gates on encrypted qubits \cite{fisher2014quantum,Broadbent2015delegating}, hiding from the remote quantum server a circuit to be implemented, known as blind quantum computation \cite{Childs:BQC:QIC05,AS:BQC:IJQIC06,BFK:UBQC:FOCS09,morimae2013secure,sheng2015deterministic,BEBFZP:DBQC:S12,MF:BQC:NC12,Morimae:BQC:PRL12,DKL:BQC:PRL12,MF:BQC:PRA13,SKM:BQC:PRA13,GMMR:BQC:PRL13,MPFF:BQC:PRL13,li2014triple,HMveri2015}, and computing on encrypted qubits by multiple-round quantum communication and complicated verification mechanism \cite{dupuis2012actively,BGS:BQC:AC13}.
 We will use the idea of the typical DQC protocol on encrypted data in Ref. \citenum{fisher2014quantum}. This
protocol can allows a user whose quantum power is limited to encryption and preparing random BB84 states, to delegate the execution of any quantum computation on encrypted data to a remote quantum server with only one round of quantum communication. It offers
perfect privacy against any adversarial server, although it does not provide a method to verify the result. We briefly review the protocol in the following. More details can be found in Ref. \citenum{fisher2014quantum}.

(D1) A client encrypts her qubits  $|\phi\rangle$ with the quantum one-time pad and then sends the encrypted qubits  $|\phi\rangle_{enc}$ to a quantum server. Specifically, for each qubit $|\phi_i\rangle$, the client performs a combination of Pauli X and Z operations on it to obtain $|\phi_i\rangle_{enc}=$X$^a$Z$^b|\phi_i\rangle$, where $a$ and $b$ are randomly chosen from $\{0,1\}$ and form the key. Obviously, with the information of $a$ and $b$, the encrypted qubit $|\phi_i\rangle_{enc}$ can be decrypted by reversing the Pauli X and Z rotations.

(D2) The server implements an agreed on quantum computation $U$ on the encrypted qubits to get $U(|\phi\rangle_{enc})$. $U$ can be universal and achieved in a general quantum circuit which can be decomposed into a serial of the following operations: quantum gates in a universal gate set \{X, Z, CNOT, H, P, R\}, auxiliary qubits prepared in $|0\rangle$, and single-qubit computational basis measurements, where the Pauli gates X and Z, the two-qubit gate CNOT, the single-qubit Hadamard gate H, the single-qubit phase gate P, and the non-Clifford gate R, have the following actions: X$|j\rangle\rightarrow|j\oplus1\rangle$, Z$|j\rangle\rightarrow(-1)^j|j\rangle$, CNOT$|j\rangle|k\rangle\rightarrow|j\rangle|j\oplus k\rangle$, H$|j\rangle\rightarrow(|0\rangle+(-1)^j)/\sqrt{2}$, P$|j\rangle\rightarrow(i)^j|j\rangle$, and R$|j\rangle\rightarrow(e^{i\pi/4})^j|j\rangle$, for $j\in\{0,1\}$.  For any Clifford gate including X, Z, CNOT, H, and P on encrypted qubits,  the client does not require any additional classical or quantum resource, and only needs to know what gates are implemented to update the decryption key. But for the non-Clifford gate R on encrypted data, the client needs preparing auxiliary qubits and classical interactions to modify the decryption key.

(D3) The server returns the output state $U(|\phi\rangle_{enc})$ to the client who will obtain $U(|\phi\rangle)$ by decrypting it with the updated decryption key that she can compute.

\section*{Results}
In this section, we first describe the protocol which will be shown to be realized with current technology, and then analyze its security and compare it with other typical SQKD protocols.

\subsection*{The proposed protocol.}

We begin to present the SQKD protocol where nearly classical Bob can generate a shared key with quantum Alice, by delegating his quantum computation to a quantum server Charlie. Let $n$ and $m$ be the desired number of sifted key bits and final shared key bits, $\delta>0$, $\theta>0$, and $\sigma>0$ be certain fixed parameters, and $l$ be the transmission speed threshold of qubits which will be useful for the security of the protocol. The detailed steps of the protocol are given as follows.

\textit{The Quantum transmission phase.}

(S1) Alice prepares $N=16n(1+\delta)$ qubits at random and sends them to Bob with a speed greater than or equal to $l$. Each qubit $|\psi_i\rangle$ is one of the four states $\{|0\rangle,|1\rangle,|+\rangle=(|0\rangle+|1\rangle)/\sqrt{2},|-\rangle=(|0\rangle-|1\rangle)/\sqrt{2}\}$, where $i=1,2,\cdots,N$. Here $|\psi_i\rangle$ can been regarded as the encrypted result of another state. For example, Alice first randomly produces a state $|\phi_i\rangle$, and then applies X$^a$Z$^b$ on it to get $|\psi_i\rangle$, namely $|\psi_i\rangle=$X$^a$Z$^b|\phi_i\rangle$, where $a$ and $b$ are randomly chosen from $\{0,1\}$ and made up of the encryption key.

(S2) As each qubit $|\psi_i\rangle$ arrives, Bob randomly decides whether to discard the qubit directly or not. For the qubit $|\psi_{s_j}\rangle$ that Bob did not throw away, Bob records its position $s_j$ ($j\in\{1,2,\cdots,8n(1+\theta)\}$), transmits it to Charlie, and ask him to apply the Pauli gate $u_{s_j}$ which is randomly chosen from $\{X,Z\}$. We should note that the transmission speed of qubits should be quick enough so that Charlie or other attackers cannot distinguish Bob's choices. We assume that the qubit-transmission speed threshold during Bob's reception is $l$ for preventing attackers to learn Bob's random choices. If Bob observers the speed value is smaller than $l$, he aborts the protocol and starts a new one.

(S3) After performing the operation $u_{s_j}$ required by Bob, Charlie reflects the qubit $u_{s_j}|\psi_{s_j}\rangle$ back to Bob still at a speed no less than $l$.

(S4) For each qubit coming from Charlie, Bob chooses either throws it away, or sends it to Charlie again and asks him to measure it in the rectilinear basis $R$ or diagonal basis $D$. Bob also observes the transmission speed of qubits and then decides whether to continue.

(S5) Charlie performs corresponding measurements on the qubits and sends all the measurement results $\{b_{t_k}\}_{k=1}^{4n(1+\sigma)}$ to Bob, where $t_k\in\{s_1,s_2,\cdots,s_{8n(1+\theta)}\}$.

\textit{The Public discussion phase.}

(S6) Alice announces the basis corresponding to the state of each qubit $|\psi_i\rangle$ she prepares. For instance, if $|\psi_i\rangle\in\{|0\rangle,|1\rangle\}$, Alice announces $R$, otherwise reveals $D$.

(S7) Bob tells Alice the positions where he chose right bases and then they discard the bits in other positions. There is a high probability that at least $2n$ positions that Alice and Bob should agree. Suppose these agreed positions be indexed by $p_1,p_2,\cdots,p_{2n}$. If $|\psi_{p_k}\rangle(k\in\{1,2,\cdots,2n)$ was prepared in $|0\rangle$ or $|+\rangle$, Alice interprets the bit as 0, otherwise interprets it as 1. But for qubits in positions $p_k$, there are four cases occurring in the same probability from the perspective of Bob: (1) the Pauli gate X and measurement in the basis $R$ are applied, (2) the Pauli gate X and measurement in the basis $D$ are applied, (3) the Pauli gate Z and measurement in the basis $R$ are applied, and (4) the Pauli gate Z and measurement in the basis $D$ are applied.  For cases (1) and (4) Bob interprets the bit as $1-b_{p_k}$, and interprets it as $b_{p_k}$ for the other two cases. By this method, Alice and Bob keep $2n$ bits.

(S8) Alice and Bob publicly announce and compare $n$ bits to check for eavesdropping and Charlie's dishonesty. If the disagreements exceed an acceptable number, they abort the protocol. Otherwise, they take the remaining $n$ bits as a sifted key.

(S9) Alice and Bob perform purely classical information reconciliation and privacy amplification on the $n$-bit sifted key to obtain the final $m$-bit shared key.

\begin{figure}[ht]
\centering
\includegraphics[width=\linewidth]{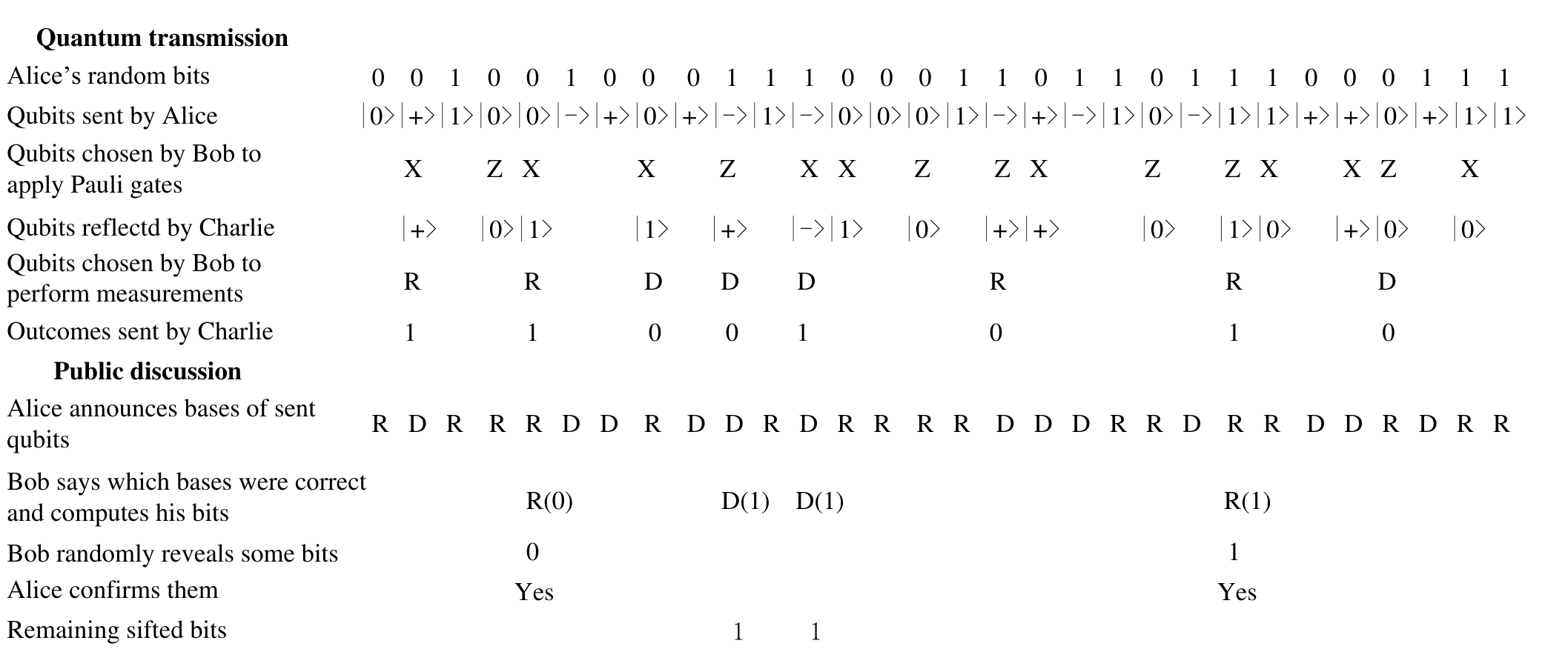}
\caption{An example for the proposed SQKD protocol.}
\label{fig:1}
\end{figure}

The above protocol can be illustrated by a specific example as shown in Fig. \ref{fig:1}. In addition, the presented protocol only needs simplified experimental requirements of quantum key distribution plus Pauli gates X and Z, which can be experimentally realized using today's technology \cite{BEBFZP:DBQC:S12}. As for the transmission speed threshold of qubits for ensuring attackers unable to know Bob's random choices, namely either to discard qubits or transmit them to the delegated server for further operations, it may be not difficult to achieve since one can currently expect at least 1.02M qubits per second for a fiber distance of 20 km and 10.1K qubits per second for 100 km \cite{dixon2008gigahertz}.

\subsection*{Security analysis.}
A SQKD protocol is usually said to be robust if for any attack of an adversary to gain information will necessarily induce some detectable errors.
We show the robustness of the proposed protocol mainly in a reduction way, with the only difference that there is an assumption on the attacker. In this protocol, attackers are not all-powerful since they are supposed to be unable to distinguish the almost classical party's random choices when a string of unknown qubits arrive.

\textbf{Secure against an eavesdropper Eve between Alice and Bob.} We first consider a special case that Eve exists only between Alice and Bob without knowing the delegated server. Then from the perspective of Eve, since nearly classical Bob can delegates all his quantum operations to Charlie for obtaining the corresponding results, the proposed protocol (\textbf{Protocol 1}) can be reduced to a protocol  (\textbf{Protocol 2}) where Alice and Bob implement a quantum key distribution protocol, similar to the famous BB84 protocol \cite{bennett1984quantum} with modifications that Bob randomly discards some qubits, or applies Pauli gates on them and measures some of them in the bases $R$ or $D$ at random. Thus \textbf{Protocol 2 }can obtain the similar level of security as the BB84 protocol, but sacrificing qubit efficiency, and so is the \textbf{Protocol 1} in this case. For example, we can suppose Eve intercepts all the qubits and measures them in the bases chosen by himself. As Eve cannot know which positions Bob chose to apply Pauli operators and perform measurements, in each position she only has a probability of 1/4 to guess the two choices right and escapes from being detected with the probability $1/4+1/4*1/2+1/4*1/2+1/4*1/2=5/8$. Then the probability that Eve goes undetected is $(5/8)^n$, compared with $(3/4)^n$ of the BB84 protocol.

\textbf{Secure against an untrusted server Charlie.} In a more general scenario, Charlie may be dishonest and also attempt to obtain some information about the shared key between Alice and Bob. We can assume there is no other third-party eavesdroppers since their attacks can be absorbed into an attack initiated by a malicious Charlie. In addition, there should be an authenticated classical channel between Alice and Bob that is normal in SQKD protocols. The classical channel from Charlie to Bob is unnecessary but better to be authenticated, since an authenticated channel can increase the successful rate of the protocol.
From the server Charlie's view of \textbf{Protocol 1}, he preforms the protocol similar to the reviewed DQC protocol \cite{fisher2014quantum} with Bob, and also can intercept and operate on all the qubits that were sent by Alice to Bob like an eavesdropper. We consider the security in two cases according to whether Charlie initiates eavesdropping on the quantum channel between Alice and Bob.

 If Charlie does not wiretap when Alice sends qubits to Bob, the security of \textbf{Protocol 1} mainly depends on the employed DQC protocol. Thus \textbf{Protocol 1} can be reduced to a modified DQC protocol, namely \textbf{Protocol 3}, which can be modeled as follows:

(D1') Alice sends to Bob a state $|\psi\rangle$ of $N$ qubits, each of which is either $|0\rangle$, $|1\rangle$, $|+\rangle$, or $|-\rangle$. The state $|\psi\rangle$ can be obtained by applying quantum one-time pad on another state $|\phi\rangle$ with two key strings K1 and K2,  namely $|\psi\rangle=E_{K_1,K_2}(|\phi\rangle)$. When Bob receives each qubit, he randomly decides to discard it or transmit it to Charlie. So the state $|\varphi\rangle$ that Charlie receives is a totally random subsystem of $|\psi\rangle$. It can be seen that Bob encrypts $|\psi\rangle$ to get $|\varphi\rangle$.

(D2') Charlie implements corresponding quantum computation $U$ on $|\varphi\rangle$ using the reviewed DQC protocol \cite{fisher2014quantum}.

(D3') Different from the step D3 in the reviewed protocol, Charlie not only returns Bob the resultant state $U|\varphi\rangle$, but also sends Bob the measurement outcomes of half qubits of $U|\varphi\rangle$ randomly chosen by Bob.

According to the security analysis in Ref. \citenum{fisher2014quantum}, Charlie cannot learn anything about $U|\psi\rangle$ and $|\psi\rangle$ from $U|\varphi\rangle$. Even if in the step D3', Charlie are required to perform measurements on some quibts, which can be regarded as that Bob asks for classical output instead of quantum output, Bob still should not find any information about $U|\psi\rangle$ and $|\psi\rangle$, otherwise the reviewed DQC protocol \cite{fisher2014quantum} cannot keep the client's data private. Thus, \textbf{Protocol 1} is as secure as \textbf{Protocol 3} before public discussion.

The process of public discussion is not only used for Alice and Bob to obtain the shared sifted key bits, but also provides a method to verify whether Charlie follows the protocol to some extent. Although Alice and Bob reveal the bases of qubits where they have the same choices, Charlie still cannot learn the bits since he does not know which qubits Bob chose Pauli X or Z operations and thus cannot know whether he should flip the measurement outcomes or not. In addition, if Charlie alters the transmission or does not perform the operations as required, extra disagreements will be induced on some of the bits that Alice and Bob think they should agree.

If Charlie controls the quantum channel from Alice to Bob, the security of \textbf{Protocol 1} does not just depend on the employed DQC protocol since the qubits that Bob receives may not be the real ones from Alice. We consider the worst case that Charlie intercepts all the qubits sent by Alice and replaces them with his own ones, such as those randomly chosen from $\{|0\rangle,$ $|1\rangle\}$ instead of $\{|0\rangle,$ $|1\rangle\}$, $|+\rangle,$ $|-\rangle\}$. Then no matter which qubits Bob chose to forward in step S2, Charlie can distinguish these orthogonal states and learn Bob's choices.  Similarly, Charlie also can figure out Bob's further choices in step S4 by measuring all the coming qubits in the same basis $R$. By doing so, Charlie can learn whatever Bob does. However, during the public discussion, for each position that Alice and Bob chose the same basis, there still has a disagreement between Alice and Bob with a probability $1/2$ since Charlie did not know the original states that Alice prepared. So the probability that Charlie is not noticed is $(1/2)^n$ which approaches zero when $n$ is big enough.

\subsection*{Comparisons.}
\begin{figure}[ht]
\centering
\includegraphics[width=\linewidth]{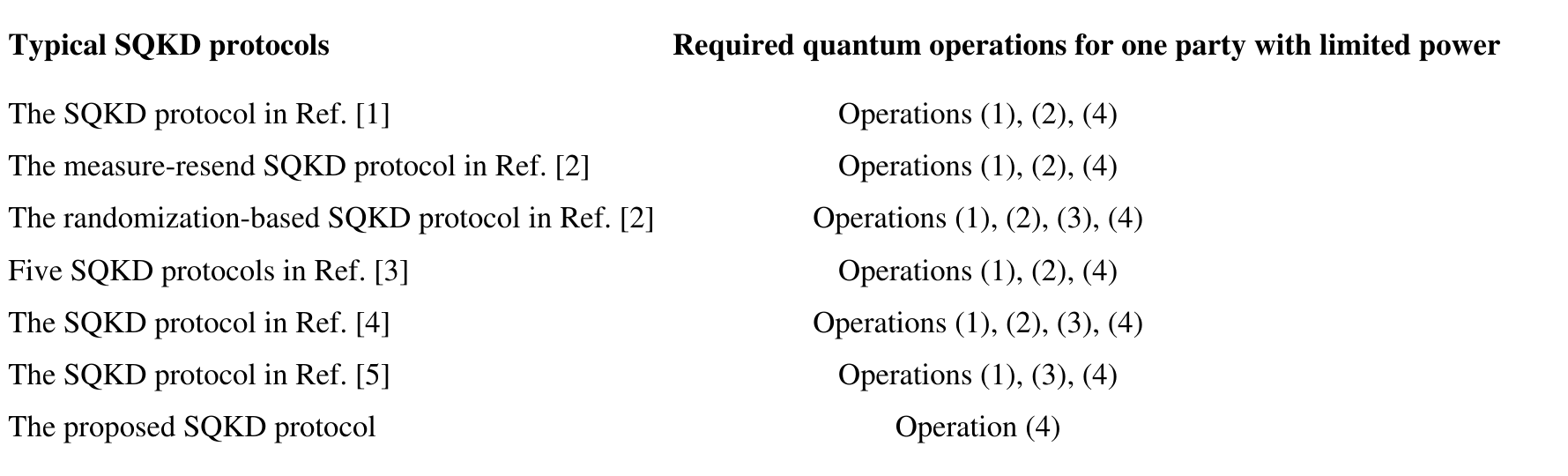}
\caption{Comparisons among several SQKD protocols.}
\label{fig:2}
\end{figure}

In existing SQKD protocols \cite{BKM:SQKD:PRL07,BGKM:SQKD:PRA09,zou2009semiquantum,jian2011semiquantum,zou2015semiquantum,yu2014authenticated,lu2008quantum,zhang2009quantum,krawec2014mediated} , one party with limited quantum power usually needs to perform three or four of the following quantum operations: (1) prepare qubits in the computational basis $\{|0\rangle,|1\rangle\}$, (2) measure qubits in the computational basis $\{|0\rangle,|1\rangle\}$, (3) reorder qubits, and (4) access quantum channels, while in the proposed protocol, the party needs to implement only operation (4). In other words, compared with the related work, our main contribution is that the quantum requirement that one party should have the ability of preparing and measuring qubits in the computational basis, or reordering qubits in typical SQKD protocols is removed and thus such party is more classical. The detailed comparisons between the given protocol and some typical ones are shown in Fig. \ref{fig:2}.

\section*{Discussion} We have proposed a SQKD protocol by employing secure DQC where almost classical Bob who does not require quantum capability or quantum memory and only needs to access quantum channels can establish a shared key with quantum Alice. The quantum resources of one party in our protocol is restricted to the minimum, so more users will have chances to participate quantum key distribution and enjoy its advantage. We also have provided an application of the DQC protocol on encrypted data recently presented in Ref. \citenum{fisher2014quantum} and offered a verification method for it to some extent. Furthermore, this is the first time to build a bridge between QKD and DQC, the combination of which will play a significant role in the advancement of secure distributed quantum applications, and throw lights on designing future quantum hybrid networks where quantum cryptographic communication and quantum computation are to be implemented.

However, we have to achieve this more practical SQKD protocol at the cost of sacrificing qubit efficiency which is only 12.5\%, compared with 25\% of the typical SQKD protocol \cite{BKM:SQKD:PRL07}. It can be significantly improved if relaxing quantum requirements of the party with restricted power, such as allowing him to have memory for reordering qubits, but quantum memory is not an easy task with current technology. How to increase the key rate in the proposed SQKD protocol will be the future work.


\begin{thebibliography}{10}
\expandafter\ifx\csname url\endcsname\relax
  \def\url#1{\texttt{#1}}\fi
\expandafter\ifx\csname urlprefix\endcsname\relax\def\urlprefix{URL }\fi
\providecommand{\bibinfo}[2]{#2}
\providecommand{\eprint}[2][]{\url{#2}}

\bibitem{BKM:SQKD:PRL07}
\bibinfo{author}{Boyer, M.}, \bibinfo{author}{Kenigsberg, D.} \&
  \bibinfo{author}{Mor, T.}
\newblock \bibinfo{title}{Quantum key distribution with classical {B}ob}.
\newblock \emph{\bibinfo{journal}{Phys. Rev. Lett.}}
  \textbf{\bibinfo{volume}{99}}, \bibinfo{pages}{140501}
  (\bibinfo{year}{2007}).

\bibitem{BGKM:SQKD:PRA09}
\bibinfo{author}{Boyer, M.}, \bibinfo{author}{Gelles, R.},
  \bibinfo{author}{Kenigsberg, D.} \& \bibinfo{author}{Mor, T.}
\newblock \bibinfo{title}{Semiquantum key distribution}.
\newblock \emph{\bibinfo{journal}{Phys. Rev. A}} \textbf{\bibinfo{volume}{79}},
  \bibinfo{pages}{032341} (\bibinfo{year}{2009}).

\bibitem{zou2009semiquantum}
\bibinfo{author}{Zou, X.}, \bibinfo{author}{Qiu, D.}, \bibinfo{author}{Li, L.},
  \bibinfo{author}{Wu, L.} \& \bibinfo{author}{Li, L.}
\newblock \bibinfo{title}{Semiquantum-key distribution using less than four
  quantum states}.
\newblock \emph{\bibinfo{journal}{Phys. Rev. A}} \textbf{\bibinfo{volume}{79}},
  \bibinfo{pages}{052312} (\bibinfo{year}{2009}).

\bibitem{jian2011semiquantum}
\bibinfo{author}{Wang, J.}, \bibinfo{author}{Zhang, S.},
  \bibinfo{author}{Zhang, Q.} \& \bibinfo{author}{C.-J., T.}
\newblock \bibinfo{title}{Semiquantum key distribution using entangled states}.
\newblock \emph{\bibinfo{journal}{Chin. Phys. Lett.}}
  \textbf{\bibinfo{volume}{28}}, \bibinfo{pages}{100301}
  (\bibinfo{year}{2011}).

\bibitem{zou2015semiquantum}
\bibinfo{author}{Zou, X.}, \bibinfo{author}{Qiu, D.}, \bibinfo{author}{Zhang,
  S.} \& \bibinfo{author}{Mateus, P.}
\newblock \bibinfo{title}{Semiquantum key distribution without invoking the
  classical party's measurement capability}.
\newblock \emph{\bibinfo{journal}{Quantum Inf. Process.}}
  \textbf{\bibinfo{volume}{28}}, \bibinfo{pages}{2981--2996}
  (\bibinfo{year}{2015}).

\bibitem{yu2014authenticated}
\bibinfo{author}{Yu, K.-F.}, \bibinfo{author}{Yang, C.-W.},
  \bibinfo{author}{Liao, C.-H.} \& \bibinfo{author}{Hwang, T.}
\newblock \bibinfo{title}{Authenticated semi-quantum key distribution protocol
  using {B}ell states}.
\newblock \emph{\bibinfo{journal}{Quantum Inf. Process.}}
  \textbf{\bibinfo{volume}{13}}, \bibinfo{pages}{1457--1465}
  (\bibinfo{year}{2014}).

\bibitem{lu2008quantum}
\bibinfo{author}{Lu, H.} \& \bibinfo{author}{Cai, Q.-Y.}
\newblock \bibinfo{title}{Quantum key distribution with classical {A}lice}.
\newblock \emph{\bibinfo{journal}{Int. J. Quantum Inf.}}
  \textbf{\bibinfo{volume}{6}}, \bibinfo{pages}{1195--1202}
  (\bibinfo{year}{2008}).

\bibitem{zhang2009quantum}
\bibinfo{author}{Zhang, X.-Z.}, \bibinfo{author}{Gong, W.-G.},
  \bibinfo{author}{Tan, Y.-G.}, \bibinfo{author}{Ren, Z.-Z.} \&
  \bibinfo{author}{Guo, X.-T.}
\newblock \bibinfo{title}{Quantum key distribution series network protocol with
  m-classical {B}obs}.
\newblock \emph{\bibinfo{journal}{Chin. Phys. B}}
  \textbf{\bibinfo{volume}{18}}, \bibinfo{pages}{2143--2148}
  (\bibinfo{year}{2009}).

\bibitem{krawec2014mediated}
\bibinfo{author}{Krawec, W.~O.}
\newblock \bibinfo{title}{Mediated semi-quantum key distribution}.
\newblock \emph{\bibinfo{journal}{Phys. Rev. A}} \textbf{\bibinfo{volume}{91}},
  \bibinfo{pages}{032323} (\bibinfo{year}{2015}).

\bibitem{li2013quantum}
\bibinfo{author}{Li, L.}, \bibinfo{author}{Qiu, D.} \& \bibinfo{author}{Mateus,
  P.}
\newblock \bibinfo{title}{Quantum secret sharing with classical {B}obs}.
\newblock \emph{\bibinfo{journal}{J. Phys. A: Math. Theor.}}
  \textbf{\bibinfo{volume}{46}}, \bibinfo{pages}{045304}
  (\bibinfo{year}{2013}).

\bibitem{wang2012semiquantum}
\bibinfo{author}{Wang, J.}, \bibinfo{author}{Zhang, S.},
  \bibinfo{author}{Zhang, Q.} \& \bibinfo{author}{Tang, C.-J.}
\newblock \bibinfo{title}{Semiquantum secret sharing using two-particle
  entangled state}.
\newblock \emph{\bibinfo{journal}{Int. J. Quantum Inf.}}
  \textbf{\bibinfo{volume}{10}}, \bibinfo{pages}{1250050}
  (\bibinfo{year}{2012}).

\bibitem{li2010semiquantum}
\bibinfo{author}{Li, Q.}, \bibinfo{author}{Chan, W.} \& \bibinfo{author}{Long,
  D.-Y.}
\newblock \bibinfo{title}{Semiquantum secret sharing using entangled states}.
\newblock \emph{\bibinfo{journal}{Phys. Rev. A}} \textbf{\bibinfo{volume}{82}},
  \bibinfo{pages}{022303} (\bibinfo{year}{2010}).

\bibitem{nie2013semi}
\bibinfo{author}{Nie, Y.-y.}, \bibinfo{author}{Li, Y.-h.} \&
  \bibinfo{author}{Wang, Z.-s.}
\newblock \bibinfo{title}{Semi-quantum information splitting using {GHZ}-type
  states}.
\newblock \emph{\bibinfo{journal}{Quantum Inform. Process.}}
  \textbf{\bibinfo{volume}{12}}, \bibinfo{pages}{437--448}
  (\bibinfo{year}{2013}).

\bibitem{zou2014three}
\bibinfo{author}{Zou, X.} \& \bibinfo{author}{Qiu, D.}
\newblock \bibinfo{title}{Three-step semiquantum secure direct communication
  protocol}.
\newblock \emph{\bibinfo{journal}{Sci. China Ser. G: Phys. Mech. Astron.}}
  \textbf{\bibinfo{volume}{57}}, \bibinfo{pages}{1696--1702}
  (\bibinfo{year}{2014}).

\bibitem{fisher2014quantum}
\bibinfo{author}{Fisher, K.} \emph{et~al.}
\newblock \bibinfo{title}{Quantum computing on encrypted data}.
\newblock \emph{\bibinfo{journal}{Nat. commun.}} \textbf{\bibinfo{volume}{5}},
  \bibinfo{pages}{3074} (\bibinfo{year}{2014}).

\bibitem{Broadbent2015delegating}
\bibinfo{author}{Broadbent, A.}
\newblock \bibinfo{title}{Delegating private quantum computations}.
\newblock \emph{\bibinfo{journal}{Canad. J. Phys.}}
  \textbf{\bibinfo{volume}{93}}, \bibinfo{pages}{941--946}
  (\bibinfo{year}{2015}).

\bibitem{Childs:BQC:QIC05}
\bibinfo{author}{Childs, A.~M.}
\newblock \bibinfo{title}{Secure assisted quantum computation}.
\newblock \emph{\bibinfo{journal}{Quantum Inf. Comput.}}
  \textbf{\bibinfo{volume}{5}}, \bibinfo{pages}{456--466}
  (\bibinfo{year}{2005}).

\bibitem{AS:BQC:IJQIC06}
\bibinfo{author}{Arrighi, P.} \& \bibinfo{author}{Salvail, L.}
\newblock \bibinfo{title}{Blind quantum computation}.
\newblock \emph{\bibinfo{journal}{Int. J. Quantum Inf.}}
  \textbf{\bibinfo{volume}{4}}, \bibinfo{pages}{883--898}
  (\bibinfo{year}{2006}).

\bibitem{BFK:UBQC:FOCS09}
\bibinfo{author}{Broadbent, A.}, \bibinfo{author}{Fitzsimons, J.} \&
  \bibinfo{author}{Kashefi, E.}
\newblock \bibinfo{title}{Universal blind quantum computation}.
\newblock In \emph{\bibinfo{booktitle}{Proceedings of the 50th Annual IEEE
  Symposium on Foundations of Computer Science}}, \bibinfo{pages}{517--526}
  (\bibinfo{year}{2009}).

\bibitem{morimae2013secure}
\bibinfo{author}{Morimae, T.} \& \bibinfo{author}{Fujii, K.}
\newblock \bibinfo{title}{Secure entanglement distillation for double-server
  blind quantum computation}.
\newblock \emph{\bibinfo{journal}{Phys. Rev. lett.}}
  \textbf{\bibinfo{volume}{111}}, \bibinfo{pages}{020502}
  (\bibinfo{year}{2013}).

\bibitem{sheng2015deterministic}
\bibinfo{author}{Sheng, Y.-B.} \& \bibinfo{author}{Zhou, L.}
\newblock \bibinfo{title}{Deterministic entanglement distillation for secure
  double-server blind quantum computation}.
\newblock \emph{\bibinfo{journal}{Sci. Rep.}} \textbf{\bibinfo{volume}{5}},
  \bibinfo{pages}{7815} (\bibinfo{year}{2015}).

\bibitem{BEBFZP:DBQC:S12}
\bibinfo{author}{Barz, S.} \emph{et~al.}
\newblock \bibinfo{title}{Demonstration of blind quantum computing}.
\newblock \emph{\bibinfo{journal}{Science}} \textbf{\bibinfo{volume}{335}},
  \bibinfo{pages}{303--308} (\bibinfo{year}{2012}).

\bibitem{MF:BQC:NC12}
\bibinfo{author}{Morimae, T.} \& \bibinfo{author}{Fujii, K.}
\newblock \bibinfo{title}{Blind topological measurement-based quantum
  computation}.
\newblock \emph{\bibinfo{journal}{Nat. Commun.}} \textbf{\bibinfo{volume}{3}},
  \bibinfo{pages}{1036} (\bibinfo{year}{2012}).

\bibitem{Morimae:BQC:PRL12}
\bibinfo{author}{Morimae, T.}
\newblock \bibinfo{title}{Continuous-variable blind quantum computation}.
\newblock \emph{\bibinfo{journal}{Phys. Rev. Lett.}}
  \textbf{\bibinfo{volume}{109}}, \bibinfo{pages}{230502}
  (\bibinfo{year}{2012}).

\bibitem{DKL:BQC:PRL12}
\bibinfo{author}{Dunjko, V.}, \bibinfo{author}{Kashefi, E.} \&
  \bibinfo{author}{Leverrier, A.}
\newblock \bibinfo{title}{Blind quantum computing with weak coherent pulses}.
\newblock \emph{\bibinfo{journal}{Phys. Rev. Lett.}}
  \textbf{\bibinfo{volume}{108}}, \bibinfo{pages}{200502}
  (\bibinfo{year}{2012}).

\bibitem{MF:BQC:PRA13}
\bibinfo{author}{Morimae, T.} \& \bibinfo{author}{Fujii, K.}
\newblock \bibinfo{title}{Blind quantum computation protocol in which {A}lice
  only makes measurements}.
\newblock \emph{\bibinfo{journal}{Phys. Rev. A}} \textbf{\bibinfo{volume}{87}},
  \bibinfo{pages}{050301(R)} (\bibinfo{year}{2013}).

\bibitem{SKM:BQC:PRA13}
\bibinfo{author}{Sueki, T.}, \bibinfo{author}{Koshiba, T.} \&
  \bibinfo{author}{Morimae, T.}
\newblock \bibinfo{title}{Ancilla-driven universal blind quantum computation}.
\newblock \emph{\bibinfo{journal}{Phys. Rev. A}} \textbf{\bibinfo{volume}{87}},
  \bibinfo{pages}{060301(R)} (\bibinfo{year}{2013}).

\bibitem{GMMR:BQC:PRL13}
\bibinfo{author}{Giovannetti, V.}, \bibinfo{author}{Maccone, L.},
  \bibinfo{author}{Morimae, T.} \& \bibinfo{author}{Rudolph, T.}
\newblock \bibinfo{title}{Efficient universal blind quantum computation}.
\newblock \emph{\bibinfo{journal}{Phys. Rev. Lett.}}
  \textbf{\bibinfo{volume}{111}}, \bibinfo{pages}{230501}
  (\bibinfo{year}{2013}).

\bibitem{MPFF:BQC:PRL13}
\bibinfo{author}{Mantri, A.}, \bibinfo{author}{Perez-Delgado, C.} \&
  \bibinfo{author}{Fitzsimons, J.}
\newblock \bibinfo{title}{Optimal blind quantum computation}.
\newblock \emph{\bibinfo{journal}{Phys. Rev. Lett.}}
  \textbf{\bibinfo{volume}{111}}, \bibinfo{pages}{230502}
  (\bibinfo{year}{2013}).

\bibitem{li2014triple}
\bibinfo{author}{Li, Q.}, \bibinfo{author}{Chan, W.}, \bibinfo{author}{Wu, C.}
  \& \bibinfo{author}{Wen, Z.}
\newblock \bibinfo{title}{Triple-server blind quantum computation using
  entanglement swapping}.
\newblock \emph{\bibinfo{journal}{Phys. Rev. A}} \textbf{\bibinfo{volume}{89}},
  \bibinfo{pages}{040302(R)} (\bibinfo{year}{2014}).

\bibitem{HMveri2015}
\bibinfo{author}{Hayashi, M.} \& \bibinfo{author}{Morimae, T.}
\newblock \bibinfo{title}{Verifiable measurement-only blind quantum computing
  with stabilizer testing}.
\newblock \emph{\bibinfo{journal}{Phys. Rev. Lett.}}
  \textbf{\bibinfo{volume}{115}}, \bibinfo{pages}{220502}
  (\bibinfo{year}{2015}).

\bibitem{dupuis2012actively}
\bibinfo{author}{Dupuis, F.}, \bibinfo{author}{Nielsen, J.} \&
  \bibinfo{author}{Salvail, L.}
\newblock \bibinfo{title}{Actively secure two-party evaluation of any quantum
  operation}.
\newblock In \emph{\bibinfo{booktitle}{Advances in Cryptology, Proceedings of
  Crypto 2012}}, \bibinfo{pages}{794--811} (\bibinfo{publisher}{Springer},
  \bibinfo{year}{2012}).

\bibitem{BGS:BQC:AC13}
\bibinfo{author}{Broadbent, A.}, \bibinfo{author}{Gutoski, G.} \&
  \bibinfo{author}{Stebila, D.}
\newblock \bibinfo{title}{Quantum one-time programs}.
\newblock In \emph{\bibinfo{booktitle}{Advances in Cryptology, Proceedings of
  Crypto 2013}}, \bibinfo{pages}{344--360} (\bibinfo{publisher}{Springer},
  \bibinfo{year}{2013}).

\bibitem{dixon2008gigahertz}
\bibinfo{author}{Dixon, A.}, \bibinfo{author}{Yuan, Z.},
  \bibinfo{author}{Dynes, J.}, \bibinfo{author}{Sharpe, A.} \&
  \bibinfo{author}{Shields, A.}
\newblock \bibinfo{title}{Gigahertz decoy quantum key distribution with 1
  {M}bit/s secure key rate}.
\newblock \emph{\bibinfo{journal}{Opt. express}} \textbf{\bibinfo{volume}{16}},
  \bibinfo{pages}{18790--18797} (\bibinfo{year}{2008}).

\bibitem{bennett1984quantum}
\bibinfo{author}{Bennett, C.} \& \bibinfo{author}{Brassard, G.}
\newblock \bibinfo{title}{Quantum cryptography: {P}ublic key distribution and
  coin tossing}.
\newblock In \emph{\bibinfo{booktitle}{Proceedings of IEEE International
  Conference on Computers, Systems and Signal Processing}},
  \bibinfo{pages}{175--179} (\bibinfo{year}{1984}).

\end{thebibliography}

\section*{Acknowledgements}

We would like to thank the editor and anonymous reviewers for helpful
suggestions and comments. This work was sponsored by the Natural Science Foundation of China (Grant No. 61202398), and the Research Grants Council of Hong Kong (Grant Nos. 18300215 and 419413).

\section*{Author contributions statement}

Q.L. devised the protocol, Q.L. and W.H.C. wrote the main manuscript, and S.Z analysed the results. All authors reviewed the manuscript.

\section*{Additional information}

\textbf{Competing financial interests:} The authors declare no competing financial interests.

\end{document}